\documentclass[prd,superscriptaddress,floatfix,nofootinbib,twocolumn,aps]{revtex4-1}
\usepackage{graphicx}
\usepackage{dcolumn}
\usepackage{bm}
\usepackage{xcolor}
\usepackage{physics}
\usepackage{amsmath,amssymb,txfonts}
\usepackage{units}
\usepackage{soul}
\usepackage[english]{babel}
\usepackage[colorlinks,linkcolor=red,citecolor=blue,urlcolor=red]{hyperref}
\renewcommand{\t}[1]{\mathrm{#1}}

\begin{document}

\title{Searching for vector dark matter with an optomechanical accelerometer}
\author{Jack Manley}
\affiliation{Department of Electrical and Computer Engineering, University of Delaware, Newark, DE 19716, USA}
\author{Mitul Dey Chowdhury}
\affiliation{Wyant College of Optical Sciences, University of Arizona, Tucson, AZ 85721, USA}
\author{Daniel Grin}
\affiliation{Department of Physics and Astronomy, Haverford College, Haverford, PA 19041, USA}
\author{Swati Singh}
\affiliation{Department of Electrical and Computer Engineering, University of Delaware, Newark, DE 19716, USA}
\email{swatis@udel.edu}
\author{Dalziel J. Wilson}
\affiliation{Wyant College of Optical Sciences, University of Arizona, Tucson, AZ 85721, USA}

\date{\today}

\begin{abstract}

We consider using optomechanical accelerometers as resonant detectors for ultralight dark matter.  As a concrete example, we describe a detector
based on a silicon nitride membrane fixed to a beryllium mirror, forming an optical cavity.  The use of different materials gives access to forces proportional to baryon (B) and lepton (L) charge, which are believed to be coupling channels for vector dark matter particles (``dark photons"). The cavity meanwhile provides access to quantum-limited displacement measurements.  For a centimeter-scale membrane pre-cooled to 10 mK, we argue that sensitivity to vector B-L dark matter can exceed that of the E\"{o}t-Wash experiment in integration times of minutes, over a fractional bandwidth of $\sim 0.1\%$ near 10 kHz (corresponding to a particle mass of $10^{-10}~\text{eV/c\textsuperscript{2}}$).  Our analysis can be translated to alternative systems such as levitated particles, and suggests the possibility of a new generation of table-top experiments.

\end{abstract}

\maketitle
The absence of evidence for dark matter's most popular candidates---WIMPS, axions, and sterile neutrinos---has led to a ``growing sense of crisis" in the astronomy community \cite{bertone2018new}.  Unlike gravitational waves, whose recent detection \cite{abbott2016observation} was the culmination of decades of focused effort, the challenge of detecting dark matter (DM) remains complicated by a basic uncertainty of what to look for. (For example, the mass of DM particles/objects remains unknown to within 90 orders of magnitude.) In response to this crisis, a growing consensus is advocating a rethinking of DM candidates and the development of a more comprehensive experimental approach \cite{bertone2018new}.

Mechanical DM detectors are of interest for two reasons.  First, 
various models predict that DM produces a force on standard model (SM) particles, for example, a strain due to coupling to fundamental constants \cite{damour2010equivalence,Arvanitaki:2015iga,manley2019searching}. 
Second, advances in the field of cavity optomechanics---largely driven by gravitational wave (GW) astronomy---have seen the birth of a new field of quantum optomechanics, in which high-$Q$ mechanical resonators are probed at the quantum limit using laser fields \cite{aspelmeyer2014cavity}.  
This has given access to exquisite force sensitivities over a range of frequencies (1 kHz - 10 GHz) which is relatively unexplored, but well-motivated, in the search for DM, corresponding to wave-like ``ultralight" DM (ULDM).

Here we consider searching for ULDM with optomechanical accelerometers, a technology being pursued in a diversity of platforms ranging from levitated microspheres to whispering gallery mode resonators \cite{monteiro2017optical,guzman2014high,li2018characterization,krause2012high}. The concept of accelerometer-based ULDM detection is also well-established \cite{Graham:2015ifn,Carney:2019cio,pierce2018searching}, \textcolor{black}{forming the basis for} searches based on GW inteferometer \cite{guo2019searching}, atom interferometer \cite{geraci2016sensitivity}, and precision torsion-balance experiments \cite{Graham:2015ifn}. From a theoretical viewpoint, it is motivated by the possibility---conceivable by various production mechanisms \cite{Nelson:2011sf,Arias:2012az,Graham:2015rva,Cembranos:2016ugq,Dror:2018pdh,Co:2018lka,Bastero-Gil:2018uel,Agrawal:2018vin,Nakayama:2019rhg,Nomura:2019cvc}---that ULDM is composed of a massive vector field \cite{Nelson:2011sf,Arias:2012az}, which could couple to SM through channels such as baryon (B) or baryon-minus-lepton (B-L) number. This coupling would manifest as an equivalence-principle-violating \cite{Graham:2015ifn} (material dependent) force on uniform bodies, or a differential acceleration of bodies separated by a distance comparable to the ULDM's de Broglie wavelength.

We wish to emphasize in this Letter that optomechanical accelerometers can also be operated resonantly, enabling high sensitivity at frequencies (1 - 100 kHz) where current broadband ULDM searches are limited, in a form-factor amenable to array-based detection \cite{Carney:2019cio}. As an illustration, we consider a detector based on a silicon nitride membrane fixed to a beryllium mirror, forming a Fabry-P\'{e}rot cavity (Fig. 1).  Through a combination of high mechanical $Q$, cryogenic pre-cooling, and quantum-limited displacement readout, we find that this detector can probe vector B-L and B ULDM with sensitivity rivaling the E\"{o}t-Wash \textcolor{black}{experiments} \cite{Wagner:2012ui} in an integration time of minutes, over a fractional bandwidth of $\sim0.1\%$.  Addressing challenges such as frequency tunability (to increase bandwidth) and scalability could enable these and similar optomechanical detectors to occupy a niche in the search for DM.

\begin{figure}[t]
\begin{center}
		\includegraphics[width=0.98\columnwidth]{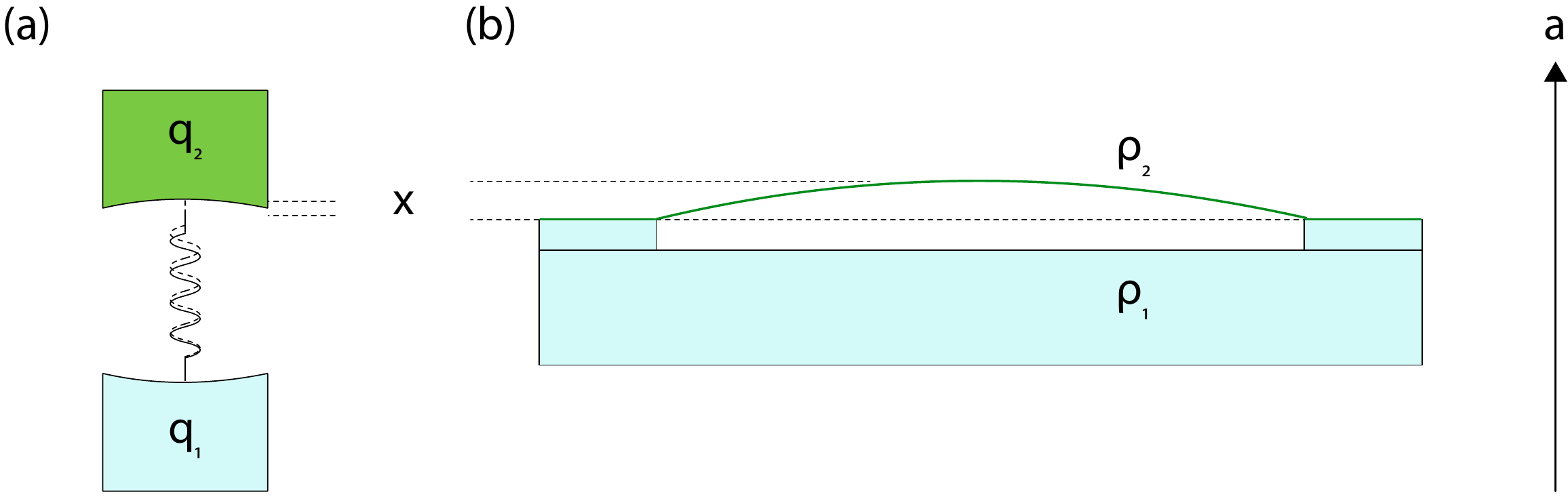}
		\caption{Concept for an optomechanical accelerometer sensitive to vector B or B-L ultralight dark matter. (a) Lumped mass model.  (b) Membrane-mirror example. Colors represent masses (materials) with different B or B-L charge (charge density), $q_i$ ($\rho_i$).  }
	\end{center}
	\vspace{-8mm}
\end{figure}

\begin{figure*}[ht!]
	\begin{center}
		\includegraphics[width=1.85\columnwidth]{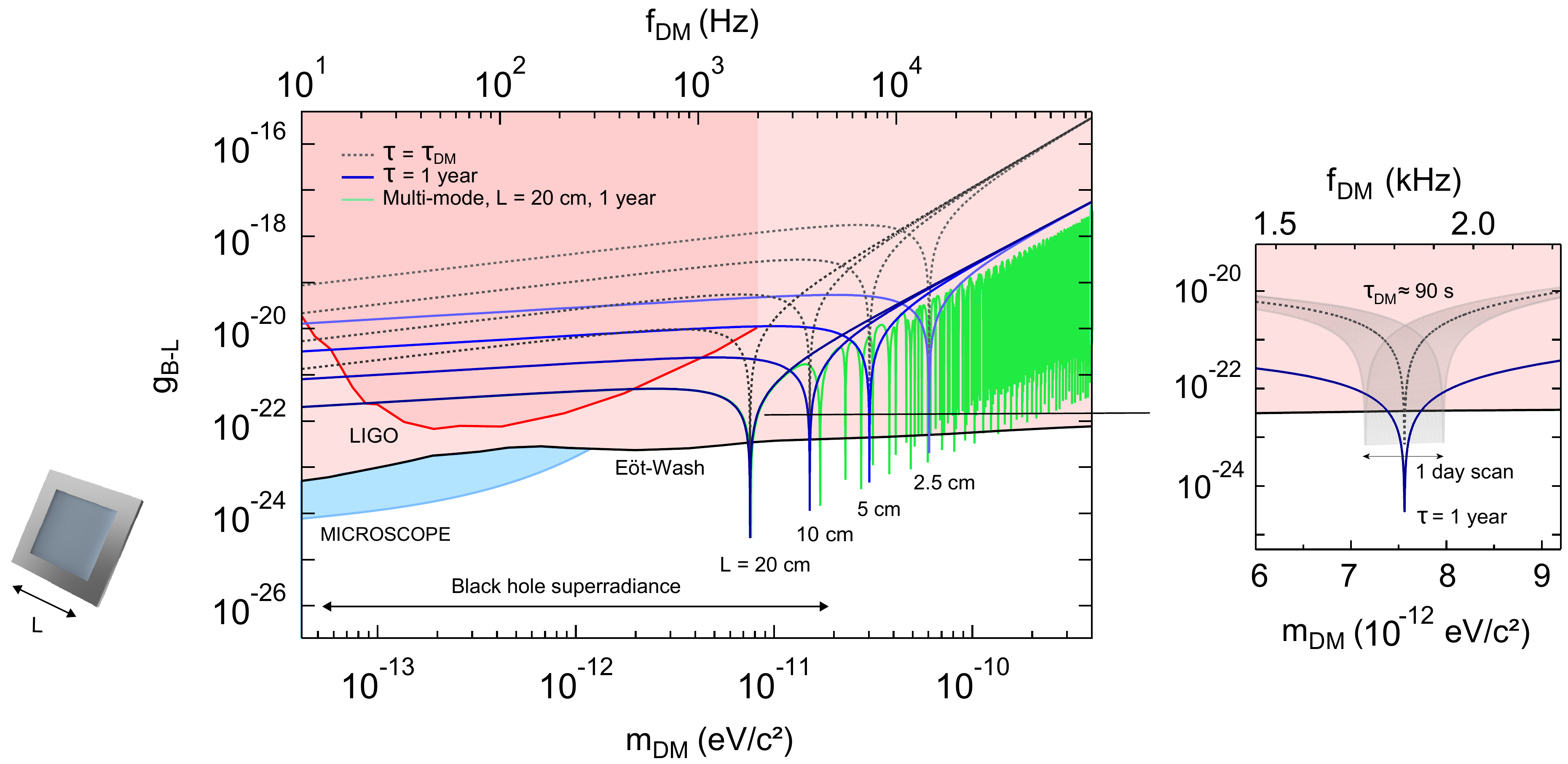}
	\caption{Centimeter-scale Si$_3$N$_4$ membranes as vector B-L dark matter detectors.  Dashed gray and solid blue curves are models for the acceleration sensitivity of four different membranes, expressed as a minimum B-L coupling strength $g_\t{B-L}$ (Eq. 12), for a measurement time equal to the DM coherence time ($\tau_\t{DM}=2Q_\t{DM}/\omega_\t{DM}$) and one year, respectively.  Each model assumes a mechanical quality factor of $Q_0=10^9$, an operating temperature of $T = 10$ mK, a displacement sensitivity of $2\times10^{-17}\,\t{m}/\sqrt{\t{Hz}}$, and a suppression factor of $f_\t{12} = 0.05$ relative to a Be reference mass.  A full multimode spectrum for the 20 cm membrane is shown in green.  Pink, red, and blue regions are bounds set by the E\"{o}t-Wash experiments, LIGO, and MICROSCOPE, respectively. At right, we zoom in on the resonance of the 20 cm membrane and illustrate a day-long scan (gray region) made in intervals $\tau_\t{DM} \approx 1.5$ min with a step size equal to the detection bandwidth $\Delta\omega_\t{det}\approx 2\pi\times 0.2$ Hz.}
	\end{center}
\vspace{-5mm}
\end{figure*}

\color{black}
Before describing the detector, it is useful to recall some basic features of ULDM. First, ultralight refers to particles of mass $m_\t{DM}\lesssim1\,\t{eV}/c^2$, which, if virialized within our Solar Neighborhood (at an average speed of $v_\t{vir}\sim 10^{-3}c$ 
\cite{freese2013colloquium}) would have a de Broglie wavelength of $\lambda_\t{DM} = h/(m_\t{DM}v_\t{vir})\gtrsim 1\,\t{mm}$. Given the local DM energy density, $\rho_{\text{DM}}\approx 0.4$ GeV/cm\textsuperscript{3}~\cite{tanabashi2018review}, the number of ULDM particles would be large with a volume $\lambda_\t{DM}^3$, implying that they behave like a coherent field.
This field would oscillate at Compton frequency $\omega_\t{DM}  = m_\t{DM}c^2/\hbar\lesssim 2\pi \times 10^{14}$ Hz with a Doppler-broadened linewidth of \textcolor{black}{$\Delta\omega_\t{DM}= \omega_\t{DM}\left(\Delta v_\t{vir}/c\right)^2\sim10^{-6}\omega_\t{DM}$}. As such, a linear detector for ULDM should look for a narrowband signal with an effective quality factor of  $Q_\t{DM}=\omega_\t{DM}/\Delta\omega_\t{DM}\sim10^6$.  Moreover, terrestrial ULDM detectors should anticipate a spatially uniform signal at frequencies $\omega_\t{DM}\lesssim2\pi\times 10\,\t{kHz}$, for which $\lambda_\t{DM}\gtrsim10^4$ km exceeds the radius of the Earth.
\color{black}

The ULDM candidates we focus on are vector (spin-1) bosons, also known as ``dark photons," coupled to B-L charge.  Composing a vector field analagous to an electromagnetic field, B-L dark photons would accelerate free-falling atoms in proportion to their charge-mass (neutron-nucleon) ratio
\begin{equation}
a(t) = g \frac{A-Z}{A} a_0\cos{(\omega_\text{DM} t +\theta_\text{DM})},
\end{equation}
where $a_0= 3.7\times10^{11}\,\text{m/s}^2$ \cite{SI}, $A\,(Z)$ is the mass (atomic) number  and $g$ is a dimensionless coupling strength. Current constraints from torsion balance equivalence principle tests (specifically, the E\"{o}t-Wash experiments \cite{Wagner:2012ui}) imply $g\lesssim10^{-22}$ for $\omega_\t{DM}\lesssim2\pi\times 10$ kHz. To exceed this bound, it is necessary to resolve accelerations at the level of
\begin{equation}
\sqrt{S_{aa}} \sim g a_0 \sqrt{\frac{Q_\t{DM}}{\omega_\t{DM}}}\lesssim 10^{-11}\sqrt{\frac{10\,\t{kHz}}{\omega_\t{DM}/2\pi}}\frac{g_0}{\sqrt{\t{Hz}}}
\end{equation}
($g_0=9.8\,\t{m}/\t{s}^2$), a task which is extreme for most accelerometers because it requires a displacement sensitivity of
\begin{equation}
\sqrt{S_{xx}} = \frac{\sqrt{S_{aa}}}{\omega_\t{DM}^{2}}\lesssim 10^{-20}\sqrt{\left(\frac{10\,\t{kHz}}{\omega_\t{DM}/2\pi}\right)^5}\frac{\t{m}}{\sqrt{\t{Hz}}}.
\end{equation}

Optomechanical accelerometers employ a mechanical resonator as a test mass and an optical cavity for displacement-based readout.
To illustrate how this can be used to detect dark photons, we first consider a lumped-mass model (Fig. 1a) in which two mirrors made of different materials, forming a cavity of length $L$$\,\ll\,$$\lambda_\t{DM}$, are attached by a massless spring.  Dark photons would produce a \emph{differential} mirror acceleration
\begin{equation}
a(t) = g f_{12}a_0 \cos{(\omega_\text{DM} t +\theta_\text{DM})},
\end{equation}
where $f_{12}$ is a purely material-dependent suppression factor
\begin{equation}
f_{12} = \abs{\frac{Z_1}{A_1}-\frac{Z_2}{A_2}}.
\end{equation}
The resulting cavity length change, $x$, which is the experimental observable, can be expressed in spectral density units as 
\begin{equation}
S_{xx}[\omega]=\frac{1}{(\omega^2-{\omega_0}^2)^2+{\omega_0}^2\omega^2/{Q_0}^2}S_{aa}[\omega],
\end{equation}
where ${\omega_0}$ and ${Q_0}$ are the frequency and quality factor of the mass-spring system, respectively.  

The advantage of the optomechanical approach is two-fold.  First, cavity enhanced readout can achieve high displacement sensitivities---an extreme case being the Laser Interferometer Gravitational-Wave Observatory (LIGO), which has achieved sensitivities of $10^{-20}\,\t{m}/\sqrt{\t{Hz}}$ at $\omega\sim2\pi\times100$ Hz, sufficient to satisfy Eq. 3~\cite{guo2019searching}.  Second, displacement sensitivity requirements are relaxed on resonance by a factor of ${Q_0}$, giving access to thermal noise limited acceleration sensitivities \cite{guzman2014high,krause2012high}
\begin{equation}
S_{aa}^\t{th} = \frac{4 k_\t{B}T\omega_\t{0}}{m Q_\t{0}}.
\end{equation}
where $T$ ($m$) is the resonator temperature (effective mass). 
 While resonant operation is not typically exploited in optomechanical accelerometers,  $\sqrt{S_{aa}^\t{th}}<10^{-11}\,g_0/\sqrt{\t{Hz}}$ might be realized in a variety of current platforms translated to cryogenic temperatures.  The trade-off, as is well known, is a ${Q_0}$-fold reduction in bandwidth, such that the sensitivity-bandwidth product is (in the ideal case of $T=0$) preserved.

Modern cavity optomechanical systems provide numerous platforms for realizing an ULDM detector. As an illustration, we consider the system sketched in Fig. 1b, consisting of a silicon nitride (Si$_3$N$_4$) membrane rigidly attached to a beryllium (Be) mirror.  It bears emphasis that special care must be taken to ensure that such an extended system is  faithful to the lumped mass model.  In particular, the use of different materials (represented by charge densities $\rho_i$ in Fig. 1b) is necessary to ensure that the suppression factor $f_\t{12}$ is non-zero ($f_{12}=0.053$ for Si$_3$N$_4$-Be). The system must also be placed in free-fall.  This can be simulated, for example, by suspending the device from a pendulum with corner frequency $\ll{\omega_0}$.

The use of a Si$_3$N$_4$ membrane is motivated by a set of features that represent the generic strengths of modern optomechanical devices, and several that make them specifically compelling for on-resonance accelerometry.  
Among these are the ability to achieve ultra-high quality factors, approaching 1 billion, using phononic engineering \cite{tsaturyan2017ultracoherent,ghadimi2018elastic}; the ability to tune resonance frequencies (to enhance bandwidth) using radiation pressure \cite{flowers2012fiber}, thermal \cite{st2019swept,sadeghi2020thermal}, and electrostatic forces \cite{naserbakht2019stress}; parts-per-million optical loss \cite{steinlechner2017optical}; and the ability to operate as a high reflectivity mirror by photonic crystal (PtC) patterning \cite{moura2018centimeter,chen2017high}.  Specific to accelerometry, a peculiar feature of membranes is their ${Q_0}$ versus mass scaling: due to an effect called dissipation dilution \cite{ghadimi2018elastic}, the ${Q_0}$ of tensily stressed membranes increases with their area, enabling large ${Q_0}\times m$ factors in a relatively compact (in one dimension) form factor.  

We thus envision, without loss of generality, a finesse $\mathcal{F}=100$ cavity formed by a Be mirror and a 200-nm-thick Si$_3$N$_4$ membrane with an embedded PtC micromirror \cite{chen2017high}. 
Probed by a coherent laser (power $P$, wavelength $\lambda$) using an ideal homodyne receiver, the detector output can be modeled as 
\begin{equation}
S_{xx}^\t{tot}=S_{xx}^\t{imp}+|\chi_{xa}(\omega)|^{2}\left(S_{aa}^\t{ba}+S_{aa}^\t{th}+S_{aa}^\t{DM}\right),
\end{equation}
where
\begin{equation}
S_{xx}^\t{imp}= \frac{\pi\hbar c\lambda}{64 \mathcal{F}^2 P}
\end{equation}
is the apparent displacement (imprecision) due to laser phase shot noise \cite{aspelmeyer2014cavity,SI},
\begin{equation}
S_{aa}^\t{ba}=\frac{\hbar^2}{m^2 S_{xx}^\t{imp}}
\end{equation}
is the acceleration (backaction) due to radiation pressure shot noise \cite{SI}, and $\chi_{xa}[\omega] = (\omega^2-{\omega_0}^2+i{\omega_0}^2/Q_0)^{-1}$ is the mechanical susceptibility (here assuming structural damping \cite{saulson1990thermal}).

In essence, ULDM detection is a parameter estimation problem. To estimate $g$, we model the dark photon signal ($a_\t{DM}$) as a Lorentzian noise peak \cite{SI}
\begin{equation}
S_{aa}^\t{DM}[\omega_\t{DM}] \approx \frac{4\langle a_\t{DM}^2\rangle}{\Delta\omega_\t{DM}}\approx\frac{2}{3}\left(\beta g f_{12} a_0\right)^2 \frac{Q_\t{DM}}{\omega_\t{DM}},
\end{equation}
here assuming a randomly polarized DM field $(\langle a^2\rangle\rightarrow\langle a^2\rangle/3$) and introducing a spatial overlap factor $\beta = (4/\pi)^2$ for the fundamental membrane mode \cite{SI}.  By feedback damping \cite{gavartin2012hybrid,harris2013minimum} or optimal filtering \cite{astone1990fast,harris2013minimum,pontin2014detection}, the narrow transfer function $|\chi_{xa}|^2$ can be inverted from the detector signal and $S_{aa}^\t{DM}[\omega_\t{DM}]$ can be estimated, for example, by averaging periodograms \cite{SI}. 
Equating $S_{aa}^\t{DM}[\omega_\t{DM}]$ with the variance in the detector noise $S_{aa}^\t{det}[\omega]=|\chi_{xa}(\omega)|^{-2}S_{xx}^\t{imp}+(S_{aa}^\t{ba}+S_{aa}^\t{th})$ yields a lower bound of \cite{budker2014proposal,SI}
\begin{subequations}
	\begin{align}
g_\t{min}[\omega] = \frac{\sqrt{3}}{\beta f_\t{12} a_0}\sqrt{\frac{ S_{aa}^\t{det}[\omega]}{\tau_\t{DM}}}&\times \begin{cases}
\left(\tfrac{\tau_\t{DM}}{\tau}\right)^{1/2} & \tau\lesssim\tau_\t{DM}\\
\left(\tfrac{\tau_\t{DM}}{\tau}\right)^{1/4} & \tau\gg\tau_\t{DM}
\end{cases}\\
\ge\frac{\sqrt{3/2}}{\beta f_\t{12} a_0}\sqrt{\frac{4k_\t{B}T\omega_\t{DM}^2}{m{Q_0} Q_\t{DM}}}&\times \begin{cases}
\left(\tfrac{\tau_\t{DM}}{\tau}\right)^{1/2} & \tau\lesssim\tau_\t{DM}\\
\left(\tfrac{\tau_\t{DM}}{\tau}\right)^{1/4} & \tau\gg\tau_\t{DM},
\end{cases}
\end{align}
\end{subequations}
\color{black}
where \color{black}$\tau_\t{DM} = 2Q_\t{DM}/\omega_\t{DM}$ \color{black} and $\tau$ are the DM coherence time and total measurement time, respectively.

\begin{figure}[t!]
	\begin{center}
		\includegraphics[width=1.0\columnwidth]{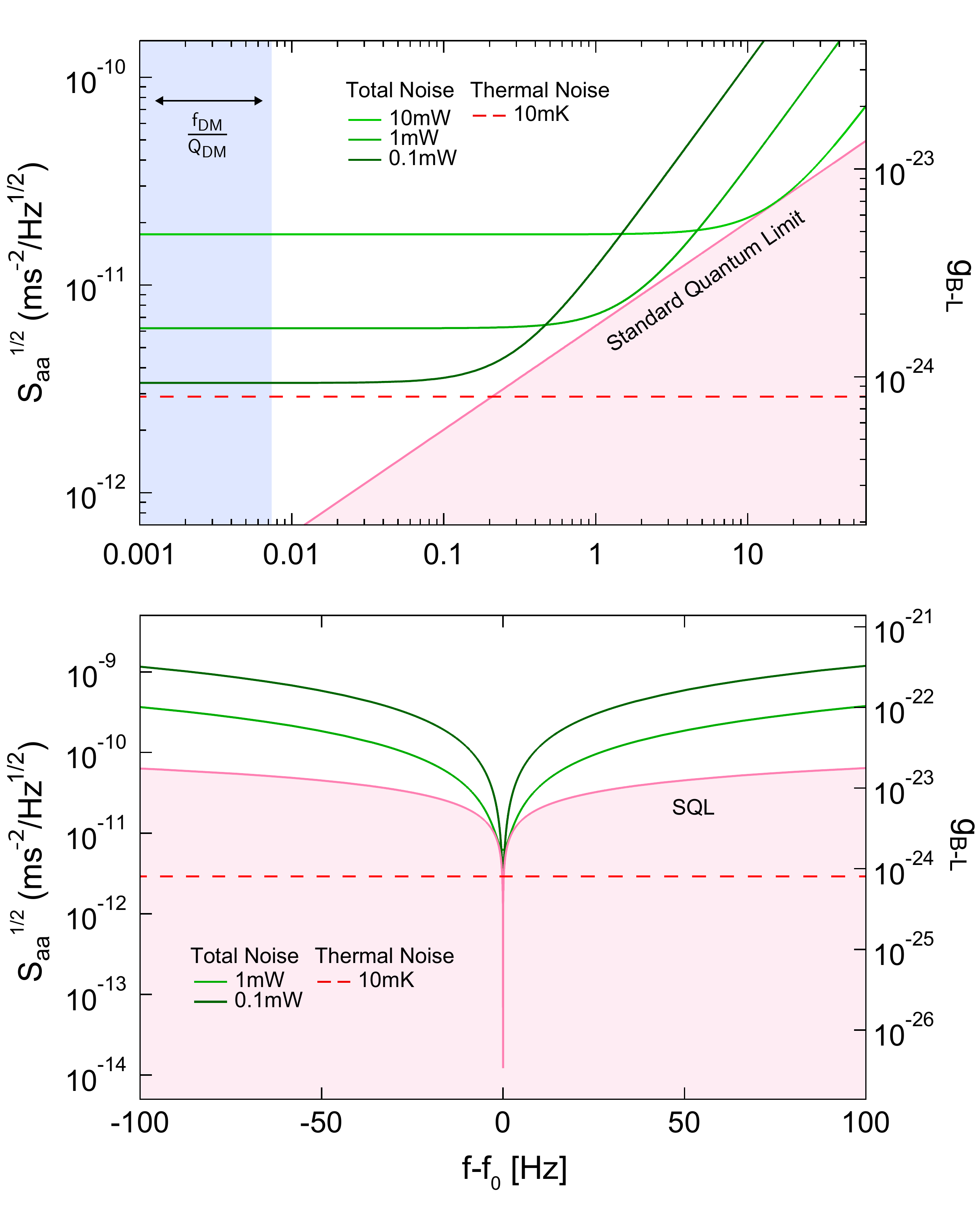}
		\caption{Detector sensitivity near resonance, versus optical power, for the 10 cm membrane in Fig. 2, indicating sensitivity-bandwidth tradeoffs due to thermal, backaction, and imprecision noise. Log-log (above) and log-linear (below) plots are shown for emphasis.  The standard quantum limit (pink line) is achieved when backaction and imprecision noise are equal, and can dominate thermal noise (dotted red line) off-resonance. The width of ULDM signal is shaded blue. }
	\end{center}
	\vspace{-5mm}
\end{figure}

In Fig. 2, we plot $g_\t{min}$ for the fundamental mode of square membranes ranging from $2.5 - 20$ cm wide, with resonance frequencies spanning from $2 - 25$ kHz.  We assume ${Q_0} = 10^9$, $T = 10$ mK, $\lambda = 1\,\mu$m, and $P =0.3$ mW, corresponding to a displacement sensitivity of $\sqrt{S_{xx}^\t{imp}}\approx 2\times 10^{-17}\,\t{m}/\sqrt{\t{Hz}}$.  We compare measurements spanning the dark matter coherence time $\tau=\tau_\t{DM}\sim10-100$ sec (dashed-gray curves) to measurements spanning one year (blue curves) and find that bounds set by the E\"{o}t-Wash experiments can be exceeded by more than an order of magnitude.  
 Comparison is made to recent constraints from LIGO in its 10 Hz - 1 kHz detection band (red curve) \cite{guo2019searching}.  Despite the large difference in displacement sensitivity between LIGO and our model system, the main difference is that LIGO's test masses are made of the same material, so that differential acceleration is produced only by the field gradient ($f_\t{12}\approx \pi L/\lambda_\t{DM}\sim 10^{-4}-10^{-6}$ \cite{SI}).

\begin{figure}[t]
	\begin{center}
		\includegraphics[width=1\columnwidth]{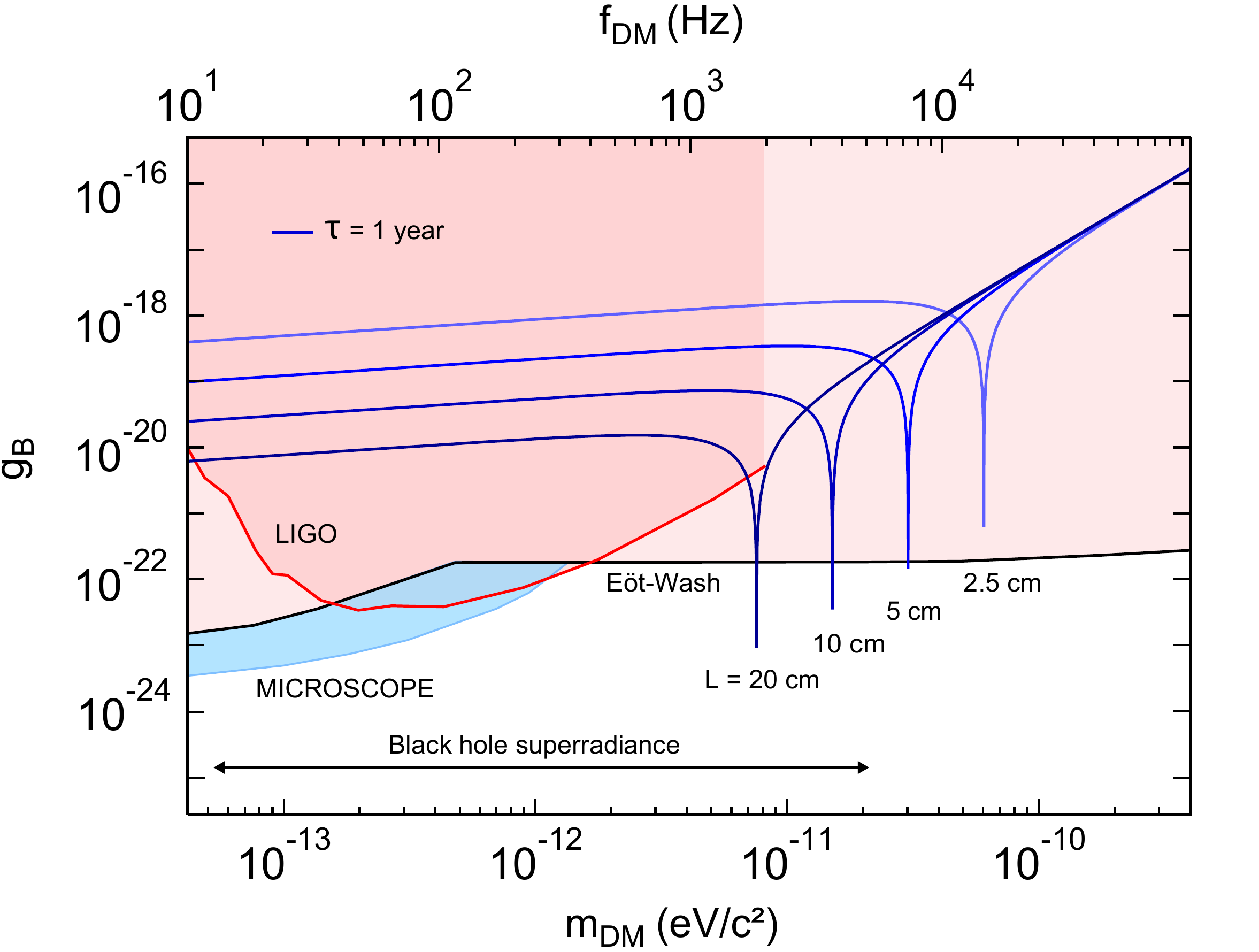}
		\caption{Projected sensitivity to B-coupled dark photons.  Blue curves are for centimeter-scale membranes as in Fig. 2, with the suppression factor $f_{12}$ adjusted for B coupling.  Black, red, and light blue curves are constraints set by the E\"{o}t-Wash experiments, LIGO, and MICROSCOPE experiments, respectively.}
	\end{center}
	\vspace{-5mm}
\end{figure}
 
 Interestingly, the bandwidth and sensitivity of our model detector is limited by quantum-back-action  \cite{Carney:2019cio}.  To visualize this tradeoff, in Fig. 3 we focus on the ``10 cm" peak in Fig. 2, and vary the optical power between 0.1 and 10 mW.  At 0.1 mW, thermal and back-action noise are equivalent, and the detection bandwidth ($\Delta\omega_\t{det}$) corresponds to the frequency range over which the total motion is resolved ($S^\t{imp}_{xx}<S_{xx}^\t{th}+S_{xx}^\t{ba}$).  At higher powers, bandwidth is increased proportionately; however, sensitivity is simultaneously reduced due to quantum back-action ($S^\t{ba}_{xx}>S_{xx}^\t{th}$).  This well known limit to the sensitivity-bandwidth product corresponds to the standard quantum limit for a force measurement, illustrated by the pink lines in Fig. 3.  In our example, evidently, maintaining sensitivity below the E\"{o}t-Wash bound ($g\sim10^{-22}$) requires limiting the fractional detection bandwidth 
 to $\Delta\omega_\t{det}/\omega_0\sim 10\%$.
 
Several techniques could improve the bandwidth of optomechanical DM detectors.
For example, a ``xylophone" detector could be realized by simultaneously monitoring multiple higher-order modes, taking advantage of the high bandwidth of optical readout. (In our case, a $L\sim1$ mm cavity length would yield a readout bandwidth of $c/(2L\mathcal{F})\sim1\t{\,GHz}$, encompassing $\sim10^{11}$ modes of a 10 cm membrane.) A full multimode spectrum \cite{SI}, shown as a green curve in Fig. 2, indicates that a single membrane can in this way yield the same performance as an array of (single mode) membranes with different sizes.  A more direct approach, similar to axion haloscope experiments \cite{braine2020extended}, would be to sweep the mechanical resonance frequency.  Considering the reduced efficiency of signal averaging for times $\tau>\tau_\t{DM}$ (Eq. 12), a natural strategy would be to step in intervals of the detection bandwidth $\Delta\omega_\t{det}$ for a total of $N=\tau/\tau_\t{DM}$ steps. For the ``10 cm" example in Fig. 2, this approach would yield an octave (a fractional bandwidth of $\Delta\omega_\t{det}N/\omega_\t{0} = 100\%$) in $\tau \sim Q_\t{DM}/\Delta\omega_\t{det}\sim$ 1 week.  Since membrane frequency scanning methods \cite{flowers2012fiber,st2019swept,sadeghi2020thermal,naserbakht2019stress} are typically limited to $\sim 10\%$ fractional bandwidth, realizing a broadband detector might in practice require a combination of xylophone, scanning, and array-based techniques.  For example, an array of $10$ membranes as shown in Fig. 2, appropriately separated in resonance frequency from 2 kHz - 4 kHz, and capable of $10\%$ fractional sweeps, could allow for an effectively broadband, thermal-noise-limited search above $\omega_\t{DM}>2\pi\times 1$ kHz, exceeding E\"{o}t-Wash bounds between 2 kHz and 20 kHz in approximately 1 week.

(In motivating such a search, it is interesting to note that $\omega_\t{DM} = 2\pi\times (1-10)$ kHz dark photons, besides having well-motivated production mechanisms \cite{Nelson:2011sf,Arias:2012az,Graham:2015rva,Cembranos:2016ugq,Dror:2018pdh,Co:2018lka,Bastero-Gil:2018uel,Agrawal:2018vin,Nakayama:2019rhg,Nomura:2019cvc}, can be independently constrained by black hole population statistics, since the corresponding Compton wavelength is comparable to the event horizon of stellar-mass black holes \cite{Brito:2015oca,Cardoso:2017kgn,Baryakhtar:2017ngi,Cardoso:2018tly}.)

Finally, we point out that the differential accelerometer approach is not limited to B-L coupling. For example, B-coupled dark photons, for which $f_\t{12}=| A_1/\mu_1- A_2/\mu_2|$ \cite{SI}, where $\mu_i$ is the mass in atomic mass units, would give rise to a differential acceleration between SiN and Be with a suppression factor of $f_{12} =0.0018$. In Fig. 4, we plot the predicted sensitivity of our detector to B-coupled dark photons compared to
the constraints set by the E\"{o}t-Wash experiments \cite{schlamminger2008test}, LIGO \cite{guo2019searching}, and MICROSCOPE \cite{touboul2017microscope,berge2018microscope}, suggesting a similar advantage at 1-10 kHz Compton frequencies.  

In summary, we have discussed the use of optomechanical accelerometers as resonant detectors for ULDM, focusing on B or B-L coupled dark photons, which produce an oscillating acceleration between masses made of different materials.  We considered an example based on a centimeter-scale Si$_3$N$_4$ membrane coupled to a Be mirror, and argued that, by combining quantum-limited displacement readout with cryogenic operating temperatures, the sensitivity of this detector can exceed current bounds, set by the E\"{o}t-Wash experiments,  in measurement time of minutes, over a fractional bandwidth of $\sim 0.1\%$ in the mass range $10^{-11}-10^{-10}$ eV/c\textsuperscript{2}. We also described scanning techniques that could broaden the bandwidth of this detector to more than an octave. Looking forward, we anticipate that a variety of optomechanical accelerometer platforms can perform similarly as vector ULDM detectors.  Optically or magnetically levitated test masses seem particularly promising, as in addition to ultra-high $Q_0\times m$ factors they can be frequency-scanned over a wide bandwidth  \cite{lewandowski2020high,monteiro2017optical,timberlake2019acceleration}. 

	We would like to thank Digesh Raut, Qaisar Shafi, Yue Zhao, Ravn Jenkins, Richard Norte, Peter Graham, and Daniel Carney for helpful discussions. This work is supported by NSF Grant No. PHY-1912480. DJW acknowledges support from a Moore Foundation Visitor Award. DG acknowledges support from the Provost's office at Haverford College.

\bibliographystyle{apsrev4-1}
\bibliography{references}
\end{document}



\title{Supplementary information for\\
	``Searching for vector dark matter with an optomechanical accelerometer"}
\author{Jack Manley}
\affiliation{Department of Electrical and Computer Engineering, University of Delaware, Newark, DE 19716, USA}

\author{Mitul Dey Chowdhury}
\affiliation{Wyant College of Optical Sciences, University of Arizona, Tucson, AZ 85721, USA}

\author{Daniel Grin}
\affiliation{Department of Physics and Astronomy, Haverford College, Haverford, PA 19041, USA}

\author{Swati Singh}
\affiliation{Department of Electrical and Computer Engineering, University of Delaware, Newark, DE 19716, USA}
\email{swatis@udel.edu}

\author{Dalziel J. Wilson}
\affiliation{Wyant College of Optical Sciences, University of Arizona, Tucson, AZ 85721, USA}

\date{\today}

\begin{abstract}

\end{abstract}
\maketitle

\numberwithin{equation}{section}

\tableofcontents
\section{Vector  B-L Dark Matter} \label{bldmappendix}
\subsection{Motivation and Production Mechanisms}
It has been hypothesized that dark matter is composed of non-thermally produced ultralight particles with $m_{\rm DM}\lesssim 10~{\rm eV/c\textsuperscript{2}}$ \cite{Marsh:2015xka}. This hypothesis is motivated by considerations from high-energy theory (in particular, a surfeit of new scalar and pseudo-scalar axion-like particles, related to the physics controlling the size of extra dimensions\cite{Arvanitaki:2009fg,Kaplan2000,deCarlos1993,Conlon:2006tq,Witten:1984dg,Svrcek:2006yi,Taylor:1988nw,Cicoli:2011wi,Damour:1994ya,Damour:2002mi,Damour:2010rp,Burgess:2010sy,Damour:1994zq,Cicoli:2012sz,Acharya:2010zx}) and astronomical measurements [indicating some tension between canonical cold dark matter (CDM) theory and small scale measurements of galaxy properties]. 

Another interesting scenario hypothesis is that DM is composed of a massive vector field, also known as a ``dark photon'' \cite{Nelson:2011sf,Arias:2012az}. Ultralight vector DM would have different (from scalars/pseudoscalars) cosmological production scenarios and gravitational imprints on the inhomogeneous universe, as discussed in Refs.~\cite{Nelson:2011sf,Arias:2012az,Graham:2015rva,Cembranos:2016ugq,Dror:2018pdh,Co:2018lka,Bastero-Gil:2018uel,Agrawal:2018vin,Nakayama:2019rhg,Nomura:2019cvc}. 

If vector DM couples to B-L, the force signature on neutron-rich elements enables direct detection, as explored in Refs. \cite{Graham:2015ifn,Carney:2019cio,pierce2018searching}. Similar effects could also be induced by a Higgs with Yukawa couplings to neutrons without equal couplings to protons and other baryons \cite{Carney:2019cio}.

There is a wide range of early-universe models for vector DM  production \cite{ Nelson:2011sf,Arias:2012az,Graham:2015rva,Dror:2018pdh,Co:2018lka,Bastero-Gil:2018uel,Agrawal:2018vin,Nakayama:2019rhg,Nomura:2019cvc}. If the vector field $A_{\mu}'$ has a non-zero vacuum expectation value in our Hubble volume, coherent oscillations could produce an adequate DM abundance \cite{Nelson:2011sf,Arias:2012az}. This requires some fine tuning due to the initial dilution of the field \cite{Arias:2012az}, motivating alternate suggestions for enhanced production from non-minimal couplings of $A'_\mu$ to the gravitational sector or SM gauge fields. 

Alternatively,~coherent oscillations of a scalar/pseudoscalar during inflation could cause resonant production of vector DM \cite{Dror:2018pdh,Co:2018lka,Bastero-Gil:2018uel,Agrawal:2018vin}. In another novel scenario, inflationary quantum fluctuations in $A'_\mu$ could generate a sufficient DM abundance, all while suppressing large-scale isocurvature fluctuations and thus evading cosmic microwave-background constraints to the ultra-light DM parameter space \cite{Graham:2015rva}.

This work considers the possibility of dark matter being entirely composed of  massive gauge bosons coupled to  B-L charge. The mass term for vector DM could in principle arise either from a novel dark Higgs sector (via spontaneous symmetry breaking) or the Stueckelberg mechanism \cite{Ruegg:2003ps,Nelson:2011sf,Arias:2012az}. The former is difficult to reconcile with constraints of adequate DM abundances and naturalness considerations \cite{Nelson:2011sf,Arias:2012az}.

\subsection{DM-Induced Acceleration of Atoms} \label{appBLforce}
In this section, we provide a brief derivation of the DM-induced acceleration signal (Eq. 1 of the main text), assuming DM to be composed of massive gauge bosons. For concreteness, we assume coupling only to B-L charge.

The Proca Lagrangian density for a massive B-L gauge field reads \cite{schambach2018proca}
\begin{equation}	\label{procalagrangian}
\mathcal{L}'=-\frac{c^2\epsilon'}{4}F'^{\mu\nu}F'_{\mu\nu}+\frac{c^2\epsilon'}{2{\lambdabar_{\rm c}}^2}A'^\nu A'_\nu-J'^\nu A'_\nu.
\end{equation}
Here, $F'_{\mu\nu}$ is the field strength tensor corresponding to the gauge field's 4-potential $A'_\mu\rightarrow \begin{pmatrix} \frac{V'}{c},	&	-\boldsymbol{A}'	 \end{pmatrix}$.  The fermionic (B-L) 4-current density is $J'_\mu$, ${\lambdabar_{\rm c}}=\frac{\hbar}{m_{\rm DM}c}$ is the gauge boson's Compton wavelength, and $\epsilon'$ is a quantity analogous to the vacuum permittivity of standard electromagnetics.

From \eqref{procalagrangian}, the field equations for the  B-L gauge field are 
\begin{equation}	\label{fieldequation}
\partial_\mu\partial^\mu A'^\nu+\frac{1}{{\lambdabar_{\rm c}}^2} A'^\nu = \frac{1}{c^2 \epsilon'}J'^\nu.
\end{equation}
Modeling dark matter as a  B-L-coupled gauge field in vacuum, \eqref{fieldequation} yields the solutions
\begin{equation}	\label{4potentialsolution}
A'^\nu=A'^\nu_0 \sin{\left(\omega t - \boldsymbol{k}\cdot\boldsymbol{r}\right)},
\end{equation}
where $\omega$ is the DM's Doppler-shifted Compton frequency, $\omega_{\rm DM}=\frac{m_\text{DM}c^2}{\hbar}$, and 
\begin{equation}
\boldsymbol{k}^2= \frac{\omega^2}{c^2}-\frac{1}{{\lambdabar_{\rm c}}^2}=\left(\frac{2\pi}{\lambda_\text{DM}}\right)^2.
\end{equation} 
The DM field oscillates in time at its Compton frequency and oscillates in space over its de Broglie wavelength $\lambda_\text{DM}=\frac{h}{m_{\rm DM}v_{\rm DM}}$.
\vspace{1mm}

We define the dark electric and dark magnetic fields analogously to electromagnetics,
\begin{equation}	\label{potentialdef1}
\boldsymbol{E}'\equiv -\nabla V' -\frac{\partial \boldsymbol{A}'}{\partial t}, \,\,\,\,\, \boldsymbol{B}'\equiv \nabla \times \boldsymbol{A}'.
\end{equation}
The force exerted on a point particle with  B-L charge takes the same form as the Lorentz force in electromagnetism. Considering a particle moving at non-relativistic speed $v_{\rm p}$ on earth, the magnetic contribution to the  B-L force due to the DM is suppressed relative to the electric contribution by a factor of $\frac{v_{\rm p}v_{\rm DM}}{c^2}$. Assuming the dark matter to have virialized, $v_{\rm DM}= v_{\rm vir}\approx 10^{-3} c$, the dark magnetic force is negligible. 

To quantify the dark electric force exerted on the particle due to DM, we introduce the coupling strength $g_\text{B-L}$, which compares the strength of the  B-L force to that of electromagnetism. We recover a force of the form
\begin{equation}	\label{DMforce2}
F\left( t, \boldsymbol{x}\right)\approx g_\text{B-L} N_\text{B-L}F_0\cos{\left(\omega t - \boldsymbol{k}\cdot\boldsymbol{r}\right)}.
\end{equation}
$N_\text{B-L}$ counts the baryon minus lepton number of the particle, and 
\begin{equation} \label{DMacceleration}
F_0\equiv \sqrt{2 \frac{e^2 \rho_{\rm DM}}{\epsilon_{0}}}\approx 6\times 10^{-16} \text{ N}
\end{equation}
is found by equating the energy density of the gauge field to the energy density of DM $\rho_{\rm DM}$.

In a displacement-sensing experiment, acceleration is a more useful quantity than force. Considering a single atom, $N_\t{B-L}=A-Z$, where $A$ ($Z$) is the mass (atomic) number. The mass is roughly $A$ times the mass of a nucleon, $m_n$, so the acceleration due to the DM force in (\ref{DMforce2}) is
\begin{equation}
a\left( t, \boldsymbol{x}\right)\approx g_\text{B-L} \frac{A-Z}{A} a_0\cos{\left(\omega t - \boldsymbol{k}\cdot\boldsymbol{r}\right)},
\end{equation}
where $a_0=\frac{F_0}{m_n}=3.7\times 10^{-11}$ m/s\textsuperscript{2}. In the main text we neglect spatial variation, and approximate $\omega\approx \omega_\t{DM}$.

\subsection{Suppression Factor}\label{appdx:Suppression}
In this section, we discuss the suppression factor $f_{12}$ that is referenced throughout the main text when comparing the DM-induced accelerations of reference and test masses.

Since an experiment is only sensitive to the differential acceleration between reference and test masses, the acceleration signal we are searching for is suppressed relative to the absolute acceleration of either mass. We use the quantity $f_{12}$ to describe this suppression.

Objects in a DM field are accelerated in proportion to their charge-to-mass ratio. In a spatially uniform field, objects composed of different materials experience a differential acceleration 
\begin{equation} \label{diffAcc}
a(t)=g f_{12} a_0 \cos{\left(\omega_\t{DM} t \right)},
\end{equation}
where the dimensionless quantity $f_{12}$ is the difference of charge-to-mass ratio (mass given in atomic mass units [amu]) between the two objects.

For coupling to B-L, the charge is $A-Z$, so 
\begin{equation}
f_{12}= \abs{\frac{A_2 -Z_2}{\mu_2}-\frac{A_1-Z_1}{\mu_1}},
\end{equation}
where $\mu_i$ is the mass in amu. Here, $f_{12}$ is primarily due to the difference in neutron fraction. In the main text, we approximate $\mu_i\approx A_i$ to get $f_{12} \approx \abs{\frac{Z_1}{A_1}-\frac{Z_2}{A_2}}$.

For coupling to B, the charge is $A$, so
\begin{equation}
f_\t{12}=\abs{\frac{A_1}{\mu_1}- \frac{A_2}{\mu_2}}.
\end{equation}
Here, $f_{12}$ is due to the difference in mass defect. 

It is worth mentioning that a differential acceleration can exist between objects of the same material. If the distance $d$ separating the objects is large enough (as for LIGO \cite{guo2019searching}), there is an appreciable phase difference in the DM field at each object; the differential acceleration effectively couples to the gradient of the DM field. To first order, considering a small phase difference $\Delta \theta_{\t{DM}}=2\pi d / \lambda_\t{DM}$, the differential acceleration between two  objects of the same material is
\begin{equation}    \label{gradAcc}
a(t)=g_\t{B-L} a_0 \left(\frac{A-Z}{A}\right)\Delta \theta_{\t{DM}} \sin{\left(\omega_\t{DM} t \right)},
\end{equation}
where we are considering coupling to B-L, for concreteness. In the main text, we simply quote $f_{12}=\left(\frac{A-Z}{A}\right)\Delta \theta_{\t{DM}}\approx \pi d/\lambda_\t{DM}$ for gradient coupling. We ignore the fact that the measurable acceleration signal is now out of phase with the DM field in contrast to our definition of $f_{12}$ in \eqref{diffAcc}), as this detail has no effect on a measurement of the signal PSD.



\subsection{DM Polarization }   \label{appdx:polarization}
Here we discuss how the unknown polarization of the DM field affects the signal strength seen by a detector.

If the DM is a vector field, it could have a preferred direction and thus polarization (with relative proportions of longitudinal and transverse modes determined by the precise production scenario \cite{ Nelson:2011sf,Arias:2012az,Graham:2015rva,Dror:2018pdh,Co:2018lka,Bastero-Gil:2018uel,Agrawal:2018vin,Nakayama:2019rhg,Nomura:2019cvc}), which will vary over a coherence time in the galactic frame \cite{Graham:2015ifn}. Here we consider this effect on a measurement taken over many coherence times. 

The maximum signal occurs when the force is parallel to the normal vector for the membrane's surface. The power in the DM signal is modified by 
\begin{equation}
S_{aa}^{\rm DM}=S_{aa,{\rm max}}^{\rm DM}\cos^2{\theta},
\end{equation}
where $\theta$ is the angle between the force and the surface normal. Over many measurements where $\theta$ is random, the average acceleration noise is
\begin{equation}
\begin{split}
\langle S_{aa}^{\rm DM}\rangle_\theta&= \frac{1}{4\pi}\int_{0}^{2\pi}  \!\!\! \int_{0}^{\pi} S_{aa,{\rm max}}^{\rm DM}\cos^2{\theta}\sin{\theta}\,{\rm d}\theta{\rm d}\phi\\
&=\frac{1}{3}S_{aa,{\rm max}}^{\rm DM}.
\end{split}
\end{equation}

\subsection{DM Lineshape Model}\label{appdx:Lineshape}
\begin{figure}[ht]
	\begin{center}
		\includegraphics[width=0.9\columnwidth]{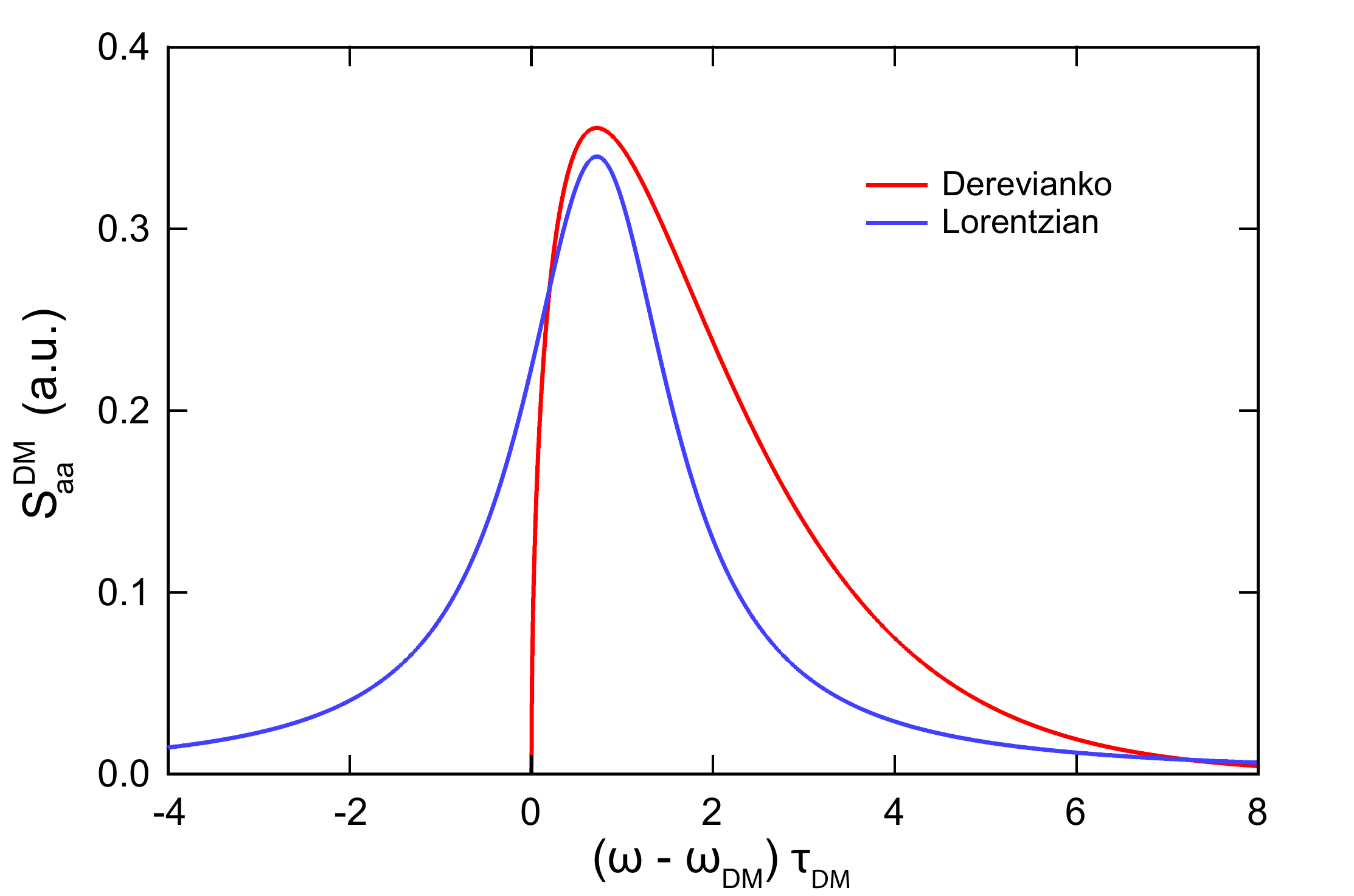}
		\caption{Comparison of DM signal lineshape. Red: PSD derived in Ref. \cite{derevianko2018detecting}. Blue: Lorentzian approximation used in this work. Here, $\omega_{\rm DM}$ is considered to be the Doppler shifted Compton frequency. Both curves are normalized such that $\int_{-\infty}^{\infty} S_{aa}^\text{DM}(\omega)=1$ in the plot.}
		\label{fig:lineshape}
	\end{center}
\end{figure}

In this section we justify modeling the DM signal PSD with a Lorentzian lineshape (Eq. 11 of the main text), which plays a role in determining how $g_\t{min}$ scales with integration time. 

The polarization-averaged, spatially uniform, monochromatic DM acceleration signal may be written as
\begin{equation}
a(t) = a_\t{DM} \cos{(\omega_\text{DM} t + \theta_\text{DM}(t))}= \frac{a_\t{DM}}{2}e^{i(\omega_\text{DM} t + \theta_\text{DM}(t))} + \text{c.c.}
\end{equation}
where $a_\t{DM} \equiv \frac{\beta g f_\t{12}}{\sqrt{3}} a_0$. $\beta$ is the spatial overlap factor -- discussed in \ref{subsec:SpatialOverlap} -- between the DM signal and the detector's vibrational modes. The factor of $\sqrt{3}$ accounts for polarization, as discussed in \ref{appdx:polarization}.

By modeling the phase $\theta_\text{DM}(t)$ appropriately, a heuristic model of the spectral shape may be obtained. We consider a random phase jump model: $\theta_\text{DM}$ remains constant up to time $t$ when it changes randomly (and uniformly) to $\left[0,2\pi\right)$. Defining the coherence time $\tau_\text{DM}$ as the average time between phase jumps, the probability of a phase change at time $t$ is given by
\begin{equation}
p(t) = \frac{1}{\tau_\text{DM}}e^{-t/\tau_\text{DM}}.
\end{equation}
Consider the auto-correlation term:
\begin{equation}
\begin{split}
G(t_\text{d}) &\equiv \langle e^{-i(\omega_\text{DM} t' + \theta_\text{DM}(t'))}  e^{i(\omega_\text{DM} (t'+t_\text{d}) + \theta_\text{DM}(t'+t_\text{d}))}\rangle_{t'}\\
&= e^{i\omega_\text{DM} t_\text{d}}\langle e^{i(\theta_\text{DM}(t'+t_\text{d}) - \theta_\text{DM}(t'))}\rangle_{t'}\\
&\equiv e^{i\omega_\text{DM} t_\text{d}}\langle e^{i\Delta\theta_\text{DM}}\rangle_{t'},
\end{split}
\end{equation}
 at time delay $t_\text{d}>0$ when averaging over times $t'$. If the phase jump occurs at time $t$ such that $t'\leq t <t'+t_\text{d}$ , 
 $\Delta\theta_\text{DM}$ is randomly distributed. Since $\Delta\theta_\text{DM}$ is a distribution,
 we replace the term $e^{i\Delta\theta_\text{DM}}$ with its expected value $\overline{e^{i\Delta\theta_\text{DM}}}$ over all possible $\Delta\theta_\text{DM}$.
 $\overline{e^{i\Delta\theta_\text{DM}}} = 0$, so the correlation $G(t_\text{d})$ is erased. 
 However, when jumps occur at $t\ge t_\text{d}$ with probability $p(t)$, $\theta_\text{DM}(t')=\theta_\text{DM}(t'+t_\text{d})$, and $e^{i\Delta\theta_\text{DM}}=1$. 
 Therefore,
\begin{equation}
\begin{split}
G(t_\text{d}) 
&= e^{i\omega_\text{DM} t_\text{d}}\int_{t_\text{d}}^{\infty}dt p(t)\\
&=e^{i\omega_\text{DM} t_\text{d}}e^{-t_\text{d}/\tau_\text{DM}}.
\end{split}
\end{equation}
This argument may be extended to negative delays ($t_\text{d}<0$) as well, giving the time auto-correlation of the acceleration signal
\begin{equation}
\begin{split}
R(t_\text{d})&\equiv\langle a(t')a(t'+t_\text{d})\rangle_{t'} \\
&=\frac{a_\t{DM}^2}{2}e^{-\abs{t_\text{d}}/\tau_\text{DM}}\frac{e^{i\omega_\text{DM}t_\text{d}}}{2}  + c.c.\\
&=\frac{a_\t{DM}^2}{2}e^{-\abs{t_\text{d}}/\tau_\text{DM}}\cos(\omega_\text{DM}t_\text{d}).
\end{split}
\end{equation}

According to the \textit{Wiener-Khinchine} theorem, the power spectral density of the signal is the Fourier transform of $R(t_\text{d})$,
\begin{equation}
\begin{split}
S_{aa}^\text{DM}(\omega)&=\int_{-\infty}^{\infty}dt_\text{d} e^{-i\omega t_\text{d}} R(t_\text{d})\\
= \frac{a_\t{DM}^2}{2}&\left[\frac{\tau_\text{DM}}{1+\tau_\text{DM}^2\left(\omega-\omega_\text{DM}\right)^2} + \frac{\tau_\text{DM}}{1+\tau_\text{DM}^2\left(\omega+\omega_\text{DM}\right)^2}\right],
\end{split}
\end{equation}
which is a \emph{double-sided} spectrum comprising both positive and negative frequencies, with two Lorentzian sidebands centered at $\omega=\pm\omega_\text{DM}$. 
In a detection experiment that cannot distinguish between positive and negative frequency components, the detected power at each frequency $\omega$ has contributions from both $S(\omega)$ and $S(-\omega)$. Allowing only non-negative frequencies, we re-define $S(\omega)$ as the detected, \emph{single-sided} PSD: the total signal power detectable in a narrow-band filter centered at $\omega\ge0$,
\begin{equation}
S(\omega\ge 0)\equiv S(\omega) + S(-\omega).
\end{equation}
Accordingly,
\begin{equation}\label{eq:AppLorSingleSided}
S_{aa}^\text{DM}(\omega\ge0)= a_\t{DM}^2\left[\frac{\tau_\text{DM}}{1+\tau_\text{DM}^2\left(\omega-\omega_\text{DM}\right)^2} \right],
\end{equation}
This is a Lorentzian with quality factor $Q_\text{DM}\equiv\frac{\omega_\text{DM}}{\text{FWHM}}=\frac{\omega_\text{DM}}{2/\tau_\text{DM}}\approx5\times10^5$ (Fig. \ref{fig:lineshape}), so, the coherence time in terms of $Q_\text{DM}$ is
$
\tau_\text{DM} = \frac{2 Q_\text{DM}}{\omega_\text{DM}} \approx \frac{10^6}{\omega_\text{DM}}
$. The DM spectrum assumes a peak signal power of
\begin{equation}
S_{aa}^\text{DM}(\omega_\text{DM})=\frac{(\beta g f_{12} a_0)^2}{3} \tau_\text{DM},
\end{equation}
from which the DM coupling strength $g$ may be inferred:
\begin{equation}\label{eq:gblSens}
\begin{split}
    g &= \frac{\sqrt{3}}{\beta f_\t{12} a_0}\sqrt{\frac{S_{aa}^\text{DM}(\omega_\text{DM})}{\tau_\t{DM}}}\\
   &= \frac{\sqrt{3/2}}{\beta f_\t{12} a_0}\sqrt{\frac{\omega_\t{DM} S_{aa}^\text{DM}(\omega_\text{DM})}{Q_\t{DM}}}.
\end{split}
\end{equation}

\section{Coupling of External Signal to Membrane Resonator}\label{appdx:ModeCoupling}
\subsection{Modes of a square, homogeneous membrane}\label{appdxsub:modes}
Let $x(y,z,t)$ the transverse displacement of the membrane (with respect to the frame) at a point $(y,z)$ of the membrane (origin at the center). $x(y,z,t)$ can be decomposed into the normal modes of the membrane ($L\times L$)
\begin{equation}
x(y,z,t)=\sum_{ij}x_{ij}(t)\phi_{ij}(y,z),
\end{equation}
where
\begin{equation}
\int dy dz \, \phi_{i'j'}(y,z)\phi_{ij}(y,z)\propto\delta_{ii'}\delta_{jj'}.
\end{equation}
The mode shape $\phi_{ij}(y,z)$ and time-dependent amplitude $x_{ij}(t)$ are determined by the 2D Helmholtz equation. Assuming uniform 2D stress $\mathcal{T}$ and uniform density $\rho$, in the absence of an external force,
\begin{subequations}
\begin{align}
\pdv[2]{x}{t}&=\left(\frac{\mathcal{T}}{\rho} \right)\laplacian x\equiv v^2\laplacian x;\\
\frac{\ddot{x_{ij}}}{v^2x_{ij}}&=\frac{\laplacian \phi_{ij}}{\phi_{ij}}=-k_{ij}^2;\\
\ddot{x}_{ij}&=-(vk_{ij})^2 x_{ij}\equiv-\omega_{ij}^2 x_{ij};\\
\laplacian \phi_{ij}&= -k_{ij}^2 \phi_{ij},
\end{align}
where we $v$ is the phase velocity in the membrane. $\omega_{ij}=2\pi f_{ij}$ are the mode frequencies corresponding to the spatial frequencies $k_{ij}$, such that $\omega_{ij}=vk_{ij}$.
\end{subequations}
The fixed boundary conditions ($\phi_{ij}(\pm L/2,\pm L/2)=0$) lead to the solutions $ \phi_{ij}(y,z)=\phi_i(y)\phi_j(z)$, where
\begin{subequations}
\begin{align}
\phi_{n}(u)=&
\begin{cases}
\cos(n\pi\frac{u}{L}) & n\in\{1,3,5,\dots\}\\
\sin(n\pi\frac{u}{L}) & n\in\{2,4,6,\dots\}
\end{cases}, \quad u\in\{y,z\}
\\
\qquad \qquad \qquad k_{ij}&=\frac{\sqrt{2}\pi}{L}\sqrt{\frac{i^2 + j^2}{2}};\\
\omega_{ij}&=vk_{ij}=\sqrt{\frac{i^2 + j^2}{2}}\omega_{11};\\
x_{ij}(t) &=x_{ij,0} \cos(\omega_{ij}t + \Phi_{ij}),
\end{align}
\end{subequations}
where $x_{ij,0}$ is the maximum displacement at the center for the $(i,j)$ mode, and $\Phi_{ij}$ is some phase depending on the choice of time origin. Typical $f_{11}\sim \left(1-10\t{kHz}\right)$ for $\sim (10\text{cm})^2$ Si\textsubscript{3}N\textsubscript{4} membranes. Note that, in theory, frequency degeneracies exist for different sets of $(i,j)$. We incorporate these degeneracies in our multi-mode sensitivity analysis (equation \ref{eq:MultiMode}); at degenerate resonances, contributions from more than one mode improves sensitivity vis-\`a-vis non-degenerate resonances (see Fig. 2 of the main text). In practice, true degeneracies are difficult to achieve in high-Q mechanical sensors due to local inhomogeneities, for example, in density and stress.

The effective mass of each mode $m_{ij}$ is defined in terms of the modal energy and amplitude:
\begin{equation}
U_{ij}\equiv m_{ij}\omega_{ij}^2 \max_{y,z}(\phi_{ij}(y,z))=m_{ij}\omega_{ij}^2.
\end{equation}
Additionally, if $m_\text{phys}=\int \rho h dydz $ denotes the physical mass of the membrane of thickness $h$, $U_{ij}$ may be written as
\begin{equation}
U_{ij}=\int dydz \ h\rho \omega_{ij}^2 \phi_{ij}^2(y,z)=\frac{m_\text{phys}}{4}\omega_{ij}^2,
\end{equation}
giving the effective modal mass
\begin{equation}
m_{ij} = \frac{m_\text{phys}}{4} \equiv m,
\end{equation}
which turns out to be mode-independent in this case.

\subsection{Mode spectrum of a spatially uniform signal}\label{subsec:SpatialOverlap}
The displacement of the center of the membrane $x(0,0,t)\equiv x_0(t)$ is measured. Only those modes $(i,j)$ with non-zero displacement at $(0,0)$ contribute to $x_0(t)$, such that there are no contributions for even $i$ or $j$. For modes that do contribute, $\phi_{ij}(0,0)=1$, and
\begin{equation}\label{totdisp}
\begin{split}
x_0(t) &=\sum_{ij}x_{ij}(t),\\
x_0(\omega) &=\sum_{ij}x_{ij}(\omega),
\end{split}
\end{equation}
where $x_0(t)$ and $x_0(\omega)$ are related by the Fourier transform.
In the presence of an external acceleration that couples to mode $(i,j)$, the equation of motion for that mode is
\begin{equation}
    \ddot{x}_{ij} + \gamma_{ij}(\omega)\dot{x}_{ij} + \omega_{ij}^2x_{ij} = a_{ij}(t).
\end{equation}
$a_{ij}$ (defined below) is the effective acceleration experienced by the individual mode, such that 
\begin{equation}\label{modeforce}
\begin{split}
x_{ij}(\omega)&=\frac{a_{ij}(\omega)}{(\omega^2-\omega_{ij}^2) + i\gamma_{ij}(\omega)\omega}\\
&\equiv
\chi_{xa,ij}(\omega)a_{ij}(\omega),
\end{split}
\end{equation}
where $\chi_{xa,ij}$ is the acceleration susceptibility of the $(i,j)$ mode. $\gamma_{ij}$ is the damping rate for each mode. We assume that the damping is mode-independent, but frequency-dependent (``structural damping"):
\begin{equation}
    \gamma_{ij}\equiv\gamma(\omega)=\frac{\omega}{Q_0}.
\end{equation}
The mechanical quality factor $Q_0$ is assumed constant across modes. 

$a_{ij}(t)$ is the fraction of the external signal that couples to a particular mode, and is given by the overlap integral between signal shape and mode shape:
\begin{equation}
    a_{ij}(t)=\frac{\int dy dz \, a(y,z,t) \phi_{ij}(y,z)}{\int dy dz \, \phi_{ij}^2(y,z)}.
\end{equation}
If $a(y,z,t)$ retains a constant shape: $a(y,z,t)=a(y,z)a(t)$, we may write
\begin{equation}
    a_{ij}(t)=\beta_{ij}a(t),
\end{equation}
where $\beta_{ij}\equiv  \frac{\int dy dz \, a(y,z) \phi_{ij}(y,z)}{\int dy dz \, \phi_{ij}^2(y,z)}$. Given a spatially uniform signal ($a(y,z)=1$, as is true locally for the DM acceleration), the spatial overlap factor for the $(i,j)$ mode is
\begin{equation}\label{eq:overlap}
\beta_{ij} = 
\begin{cases}
\left(\frac{4}{\pi}\right)^2\left(\frac{1}{ij}\right)\left(-1\right)^{\left(\frac{i+j}{2}-1\right)} & i\land j\in\{1,3,5,\dots\}\\
0 & i\lor j\in\{2,4,6,\dots\}.
\end{cases}
\end{equation}
\subsection{Multi-mode Noise}\label{appdxsub:multimode}
Phase imprecision noise $S_{xx}^\text{imp}$ (see \ref{appdx:impnoise}) is a mode-independent displacement noise. For each of the other sources of membrane displacement -- DM signal, thermal noise, and radiation pressure back-action noise -- the total detected power in each frequency bin is the sum of contributions from each independently oscillating resonator mode. The total detected PSD is
\begin{equation}
\begin{split}
    S_{xx}^\text{tot} &= S_{xx}^\text{imp} + \sum_{ij}\abs{\chi_{xa,ij}}^2\left(S_{aa}^\text{th} + S_{aa}^\text{ba} + S_{aa}^\text{DM}\right)\\
    &\equiv S_{xx}^\text{n,tot} +  \sum_{ij}\abs{\chi_{xa,ij}}^2 \beta_{ij}^2 \frac{(\beta g f_{12} a_0)^2}{3} \tau_\text{DM}\\
    &\equiv S_{xx}^\text{n,tot} + S_{xx}^\text{DM,tot},
\end{split}
\end{equation}
where $S_{xx}^\text{n,tot}$ ($S_{xx}^\text{DM,tot}$) represent the net multi-mode noise (DM) spectra.

For the DM signal to have $SNR\equiv\frac{S_{xx}^\text{DM,tot}}{\Delta S_{xx}^\text{n,tot}}=1$ (where $\Delta S^\t{n}\approx S^\t{n}$ is the standard deviation of the noise floor discussed in \ref{appdx:PdgmAvg}), it is required that $S_{xx}^\text{DM,tot} \approx S_{xx}^\text{n,tot}$. Therefore,
\begin{equation}
    g=\frac{\sqrt{3}}{\beta f_\t{12} a_0}\sqrt{\frac{S_{aa}^\text{DM,eff}(\omega_\text{DM})}{\tau_\t{DM}}},
\end{equation}
where
\begin{equation}\label{eq:MultiMode}
    S_{aa}^\text{DM,eff} \equiv \frac{S_{xx}^\text{imp} + \sum_{ij}\abs{\chi_{xa,ij}}^2\left(S_{aa}^\text{th} + S_{aa}^\text{ba} \right)}{\sum_{ij}\abs{\chi_{xa,ij}}^2 \beta_{ij}^2}
\end{equation}
is the effective noise floor after including multi-modal spatial overlaps $\beta_{ij}$.


\section{Finite Duration Estimates of Coherent Signals}\label{appdx:PdgmAvg}
The power spectrum is measured by a periodogram estimate. A periodogram obtained from a signal $a(t)$ is an average of $N$ independent measurements of $a(f)$ or $a(\omega)$ using a DFT. The periodogram $\mathcal{P}_{aa}(\omega)\Delta f\equiv\langle \abs{a(2 \pi f)}^2 \rangle_N$ over a frequency bin $\Delta f$ converges to the single-sided detected PSD in the limit 
\begin{equation}
S_{aa}(\omega)\Delta f=\lim_{N\rightarrow\infty}\mathcal{P}_{aa}(\omega)\Delta f=\lim_{N\rightarrow\infty}\langle \abs{a(2\pi f)}^2 \rangle_N.
\end{equation}
In other words, a periodogram is an accurate estimate of the true power spectrum if it is measured over an infinitely long time duration. In practice, if a signal has a finite coherence time $\tau_\text{DM}$ corresponding to a spectral feature of line-width $\Delta f_\text{DM}\sim\frac{1}{\tau_\text{DM}}$, a periodogram duration $\Delta t=\tau_\text{DM}$ is sufficiently long to resolve the PSD peak. 

A periodogram of a white noise process, $\mathcal{P}_{aa}^\t{n}(\omega)$ (whether displacement or acceleration noise), has the property that, the standard deviation in the periodogram estimate is approximately as large as the mean: $\Delta \mathcal{P}_{aa}^\t{n} \lesssim \mathcal{P}_{aa}^\t{n}$ \cite{budker2014proposal}. In order to distinguish a coherent tone from a random fluctuation of the noise floor quantified by $\Delta \mathcal{P}_{aa}^\t{n}$, the signal-to-noise, $SNR\equiv\frac{\mathcal{P}_{aa}^\t{DM}}{\Delta \mathcal{P}_{aa}^\t{n}}\approx\frac{\mathcal{P}_{aa}^\t{DM}}{\mathcal{P}_{aa}^\t{n}}$, must exceed $1$. So, the detection sensitivity of small signals $\mathcal{P}_{aa}^\t{DM}$ is improved by minimizing the noise $\mathcal{P}_{aa}^\t{n}$, which in turn minimizes the fluctuations in that noise floor.

We discuss the impact of varying integration times on signal sensitivity. Consider the two cases where the total signal observation time -- or integration time, denoted by $\tau$ -- is shorter or longer than $\tau_\text{DM}$:
\begin{enumerate}
	\item \underline{$\tau<\tau_\text{DM}$}\\
	In this regime, the frequency resolution $\Delta f\sim\frac{1}{\Delta t}$ of a single periodogram improves with increased $\Delta t$ up to $\tau_\text{DM}$. Therefore, the optimal strategy is to maximize $\Delta t$ by maximizing the periodogram duration, that is, by measuring only one periodogram of $\Delta t = \tau$.
	The power in the periodogram $\mathcal{P}_{aa}\Delta f$ measured in the frequency bin $\Delta f$ must match the total power in the true spectrum $S_{aa}\Delta f_\text{DM}$ in a bin of size $\Delta f_\text{DM}$, such that
	\begin{equation}
	\frac{S_{aa}(\omega)}{\tau_\text{DM}} = \frac{\mathcal{P}_{aa}(\omega)}{\tau}.
	\end{equation}
	Therefore, the sensitivity of the $g$ measurement in terms of a periodogram estimate of the DM signal $\mathcal{P}_{aa}^\text{DM}(\omega)$ is
	$
	g \propto \sqrt{\frac{S_{aa}^\text{DM}(\omega_\text{DM})}{\tau_\text{DM}}}
	 = \sqrt{\frac{\mathcal{P}_{aa}^\text{DM}(\omega_\text{DM})}{\tau_\text{DM}}}\sqrt{\frac{\tau_\text{DM}}{\tau}},
    $
    such that the minimum coupling measureable from an \emph{SNR} of 1 is
    \begin{equation}
       g_\t{min} = \sqrt{\frac{\mathcal{P}_{aa}^\text{n}(\omega_\text{DM})}{\tau_\text{DM}}}\sqrt{\frac{\tau_\text{DM}}{\tau}}.
    \end{equation}
	\item \underline{$\tau>\tau_\text{DM}$}\\
	A periodogram $\mathcal{P}_{aa}(\omega)$ of duration $\Delta t>\tau_\text{DM}$ samples an incoherent signal, and the signal appears flat in the frequency domain. The knowledge about a flat spectrum is not improved by improving the resolution $\frac{1}{\Delta t}$. Therefore, the periodogram duration that conveys useful information about a signal is upper bounded by the signal's coherence time $\tau_\text{DM}$.
	Since \textit{SNR} is not improved by longer $\Delta t$, the alternative is to reduce the variance in the noise floor by averaging multiple periodogram measurements, each of duration $\Delta t=\tau_\text{DM}$ over a total integration time $\tau$. \vspace{1mm} At worst, $N=\frac{\tau}{\tau_\text{DM}}$ statistically independent periodograms can be acquired. Therefore, the standard deviation in the periodogram of the acceleration noise, $\Delta P^\text{n}$, is decreased:
	\begin{equation}\label{averaging}
	\Delta \mathcal{P}_{aa}^\text{n}(\omega)=\frac{\mathcal{P}_{aa}^\text{n}(\omega)}{\sqrt{N}}=\frac{\mathcal{P}_{aa}^\text{n}(\omega)}{\sqrt{\tau/\tau_\text{DM}}}.
	\end{equation}
	This means that averaging improves upon the single-periodogram sensitivity $\sqrt{\mathcal{P}_{aa}^\text{n}}$ by a factor $N^\frac{1}{4}$, so,
	\begin{equation}\label{Tint}
	g_\t{min}\propto\sqrt{\frac{\mathcal{P}_{aa}^\text{n}(\omega_\text{DM})}{\tau_\text{DM}}}\left(\frac{\tau_\text{DM}}{\tau}\right)^\frac{1}{4}.
	\end{equation}
	Since all measurements of the true PSD $S_{aa}$ involve the periodogram estimate $\mathcal{P}_{aa}$, we re-label $\mathcal{P}_{aa}$ as $S_{aa}$ in the rest of the paper; the periodogram estimation is implied.
\end{enumerate}

\section{Displacement Imprecision Noise}\label{appdx:impnoise}
In a Fabry-Perot cavity of length $L$, free-spectral range $\omega_\text{FSR} = 2\pi\frac{c}{2L}$, finesse $\mathcal{F}=\frac{\omega_{\text{FSR}}}{\kappa}$ ($\kappa$ is the cavity's output power coupling rate), and coupling constant $G=\frac{\omega_\text{cav}}{L}$ (where $\omega_\text{cav}= q \omega_\text{FSR}, \ q\in\{1,2,3,\dots\}$ is a cavity resonance frequency), the \emph{single-sided} imprecision noise in the displacement  is given by \cite{aspelmeyer2014cavity}
\begin{equation}
S_{xx}^\text{imp}=\frac{1}{8}\frac{\kappa}{\bar{n}_\text{cav}G^2}
\end{equation}
in the limit $\kappa\gg\omega_\text{cav}$. Here, $\bar{n}_\text{cav}$ is the average number of circulating photons in the cavity. We have assumed a single-sided optical cavity and a laser field that is in resonance ($0$ detuning) with the cavity. In steady state, $\bar{n}_\text{cav}$ is related to the laser power $P$ as
\begin{equation}
\bar{n}_\text{cav} = \frac{4P}{\kappa\hbar\omega_\text{cav}}.
\end{equation}
Noting that, for wavelength $\lambda = \frac{2\pi c}{\omega_\text{cav}}$,
\begin{equation}
\begin{split}
\frac{\kappa}{G}&=\left(\frac{\omega_\text{FSR}}{\mathcal{F}} \right)\left(\frac{L}{\omega_\text{cav}} \right)\\
&= \left(2\pi\frac{c}{2L} \frac{1}{\mathcal{F}}\right) \left(\frac{L \lambda}{2\pi c}\right)\\
&= \frac{\lambda}{2\mathcal{F}}.
\end{split}
\end{equation}
The detected imprecision noise depends on the detector efficiency $0\leq\eta\leq1$ such that $P_\text{det}=\eta P$. Therefore,
\begin{equation}
\begin{split}
S_{xx}^\text{imp}&= \frac{1}{8} \frac{\hbar \omega_\text{cav}}{4P_\text{det}}\frac{\kappa^2}{G^2}\\
&= \frac{\hbar}{32}\left(\frac{2 \pi c}{\lambda}\right)\left(\frac{\lambda}{2 \mathcal{F}}\right)^2\frac{1}{\eta P}\\
&= \frac{\pi\hbar c\lambda}{64}\left(\frac{1}{\eta \mathcal{F}^2 P}\right).
\end{split}
\end{equation}
We assume $\eta = 1$.


\bibliographystyle{apsrev4-1}
\bibliography{references}